\documentclass[a4paper,fleqn,usenatbib]{mnras}

\usepackage[T1]{fontenc}
\usepackage{ae,aecompl}

\usepackage{graphicx}	
\usepackage{amsmath}	
\usepackage{amssymb}	

\newcommand{\pdif}[2]{\ensuremath{ \frac{\partial #1}{\partial #2}}} 
\newcommand{\f}[2]{\frac{#1}{#2}} 

\title[Dynamics of BLR clouds subjected to a time-varying flux]
{On the efficient acceleration of clouds\\ in active galactic nuclei}

\author[Waters \& Proga]{
Tim Waters$^{1}$\thanks{E-mail: waterst3@unlv.nevada.edu}
and
Daniel Proga$^{1}$
\\
  $^1$ Department of Physics \& Astronomy, University of Nevada, Las Vegas, 
  4505 S. Maryland Pkwy, Las Vegas, NV, 89154-4002, USA\\
}

\date{Accepted XXX. Received YYY; in original form ZZZ}

\pubyear{2016}

\begin{document}
\label{firstpage}
\pagerange{\pageref{firstpage}--\pageref{lastpage}}
\maketitle

\begin{abstract}
In the broad line region of AGN, acceleration occurs naturally when a 
cloud condenses out of the hot confining medium due to the increase in 
line opacity as the cloud cools.  However, acceleration by radiation 
pressure is not very efficient when the flux is time-independent,
unless the flow is one-dimensional.
Here we explore how acceleration is affected by a time-varying flux, 
as AGN are known to be highly variable.  
If the period of flux oscillations is longer than the thermal timescale, 
we expect the gas to cool during the low flux state, and therefore line opacity should quickly increase.  
The cloud will receive a small kick due to the increased radiation force.  
We perform hydrodynamical simulations using \textsc{Athena} to confirm this effect 
and quantify its importance.  
We find that despite the flow becoming turbulent in 2D due to hydrodynamic instabilities, 
a 20\% modulation of the flux leads to a net increase in acceleration 
--- by more than a factor of 2 --- in both 1D and 2D.  
We show that this 
is sufficient to produce the 
observed line widths, although we only consider optically thin clouds. 
We discuss the implications of our results for 
photoionization modeling and reverberation mapping.
\end{abstract}


\section{Introduction}
Broad line region (BLR) clouds are central components of unification models for
active galactic nuclei (AGN) (e.g., Antonucci 1993; Urry \& Padovani 1995)
and form the basis of many photoionization and dynamical modeling efforts 
(see Netzer 2008, 2015 for recent reviews).  
Dating as far back to the first models invoking BLR clouds 
are other studies pointing out shortcomings with this picture 
(for an overview, see e.g., Krolik 1999; Netzer 2013).  
As summarized by Rees et al. (1989), BLR clouds are 
unstable to drag and radiation pressure forces when they move
supersonically through a hot confining medium (e.g., Krolik 1977;
Mathews \& Blumenthal 1977), while the standard two-phase
model of Krolik, McKee, \& Tarter (1981) 
predicts unrealistically high temperatures 
if clouds are \emph{not} to move supersonically.  
The concern is that disruptive processes prevent clouds from
reaching the high velocities necessary to account
for the widths of broad emission lines in AGN.  

In Proga \& Waters (2015; hereafter Paper~1), 
we revisited this longstanding problem and proposed
that a newly forming cloud can be accelerated to a high velocity because as the 
thermally unstable gas condenses, its opacity increases, 
and that alone enhances the transfer of momentum from the radiation field to the cloud.  
While we demonstrated that this mechanism works in 1D,
we confirmed that clouds are indeed disrupted in 2D by  
hydrodynamical instabilities.  
Compared to 1D, the acceleration of the resulting turbulent, 
two-phase medium was more than a factor of 5 times smaller.  

Motivated by multi-wavelength observations of rapid variability (on timescales of hours to days) 
that are overall suggestive of AGN variability being intrinsic rather than absorption driven 
(see Uttley \& Casella 2014 for a review), in this paper we set out to relax our assumption of a 
constant flux and show that this can further enhance the coupling between the gas and radiation field.  
To see that a time-varying flux leads to an additional acceleration mechanism, 
note the expected asymmetric response of gas that is not fully ionized: 
during low flux states when the gas can cool, the increase in the radiation pressure force can be 
substantial because the line opacity is a sensitive function of temperature.

This letter is organized as follows.   
In \S{2} we discuss our modifications to
the methods developed in Paper~1.
In \S{3}, we present our results, verifying that this acceleration mechanism is realized,  
even in 2D.  In \S{4} we discuss the implications of these results 
and show that highly supersonic velocities are obtainable 
on time and length scales consistent with typical BLR parameters.  

\section{Methods}
Using \textsc{Athena} (Stone et al. 2008), we numerically solve the following
equations of hydrodynamics:
\begin{equation}
\pdif{\rho}{t} + \mathbf{\nabla} \cdot \left(\rho \mathbf{v} \right) = 0 \label{eq:mass} ,
\end{equation}
\begin{equation}
\pdif{\left(\rho\mathbf{v}\right)}{t} + \mathbf{\nabla} \cdot \left( \rho\mathbf{v} 
\mathbf{v} + p\,\mathbb{I}\right) = \mathbf{f}_{rad},\label{eq:mom}
\end{equation}
\begin{equation}
\pdif{E}{t} + \mathbf{\nabla} \cdot [(E + p)\mathrm{v}] =
-\rho {\mathcal L} + \kappa_{eq}\nabla^2T +  \mathbf{f}_{rad}\cdot\mathbf{v}.
\label{eq:energy}
\end{equation}
Here, $(\rho, \mathbf{v}, p)$ are the primitive fluid variables, 
$\mathbb{I}$ is the unit tensor, $\mathbf{f}_{rad}$ is 
the radiation force (see below), and 
$E = p/(\gamma-1) + \rho \mathrm{v}^2/2$ is the total energy density, 
with $\gamma = 5/3$ the adiabatic index.  
Our net cooling function $\mathcal{L}$ depends on both the temperature $T$
and the photoionization parameter $\xi = 4\pi F_{\rm{ion}}/n$, 
where $F_{\rm{ion}}$ is the ionizing flux
incident on the gas and $n = \rho/\mu m_p$ is the number density ($\mu =1$
in this work).  The function $\mathcal{L}$ accounts for Compton heating/cooling 
and losses to due to bremsstrahlung using standard analytic expressions, as well as  
X-ray heating/cooling and cooling due to line emission using fits from photoionization 
calculations given by Blondin (1994; Blondin's expressions are reproduced in Paper~1).  
These calculations assumed a $T_X = 10\,\rm{keV}/k_B$ bremsstrahlung spectrum 
incident on optically thin gas of cosmic abundances.  
We estimate the conduction coefficient $\kappa_{eq}$ using 
the value for a fully ionized plasma (Spitzer 1962) evaluated at 
the radiative equilibrium temperature $T_{eq}$, defined by $\mathcal{L}(\xi,T_{eq}) = 0$.  
 
The above equations differ from those solved in Paper~1 only by the introduction of 
an oscillating ionizing flux, 
\begin{equation}
F_{\rm{ion}} (t) = F_{X} + \Delta F_X(t) = F_{X} \left( 1 + A_X \sin(2\pi\omega_X t) \right),
\end{equation}
where $A_X$ and $\omega_X$ denote the amplitude and frequency of oscillations.  
The constant (time-averaged) ionizing flux $F_{X}$ is set once values for the photoionization 
parameter and number density characteristic of the AGN environment are chosen, i.e. 
$F_{X} =n_{eq}\, \xi_{eq} /4\pi$.  
We adopt the values $\xi_{eq} = 190 \: \rm{erg~cm~s^{-1}}$ and 
$n_{eq} = 10^{14}/T_{eq}$,
yielding $n_{eq} = 5.17 \times 10^8 \,\rm{cm^{-3}}$ and 
$F_{X} = 7.82\times10^9\,\rm{erg\, s^{-1} cm^{-2}}$.  
This choice leads to the formation of a cloud with the highest possible density 
that still permits the cloud to remain optically thin, throughout its evolution, 
to the transitions contributing to the line driving. 

The radiation field incident on the gas is approximated as consisting only 
of the time-dependent ionizing flux $F_{\rm{ion}} (t)$ and a constant 
non-ionizing flux $F_{UV}$.  This distinction is made to separate the high-energy
photons responsible for setting the heating/cooling rates from the lower energy photons
that only contribute to the radiation force (through Thomson and line scattering).  
Then the radiation force, assumed to be due to a distant source, is 
\begin{equation}
\mathbf{f}_{rad} =  \f{\rho\sigma_e}{c}\Big[(1+M_{\rm max})F_{UV} + 
(1+\sigma_X)F_{\rm{ion}}(t) \Big] \hat{r} .
\end{equation}
Here $\sigma_e$ is the mass scattering coefficient for free electrons, 
$M_{\rm max}$ is the force multiplier quantifying the enhancement of scattering opacity 
due to spectral lines, and $\sigma_X$ is an effective bound-free opacity 
(in units of $\mu m_p \sigma_e$), 
which is self-consistently determined from the X-ray heating rate (see Paper~1).  
In the optically thin limit that we explore,
the force multiplier takes on its maximum value and is just the sum of the opacity 
contributions from all spectral lines (Castor et al. 1975; Owocki et al. 1988).  
A minor technical modification to the prescription for $M_{\rm max}$ used in Paper~1 
was required in order to properly study the effects of a time-varying flux: 
we mapped the increase in line opacity from $\xi$ to $T$, 
since the primary physical dependence is on temperature, 
which maps one-to-one onto $\xi$, but only for a constant flux.  
To complete the specification of the radiation force, 
we quantify $F_{UV}$ in terms of $F_{X}$ by introducing the 
parameter $f_{UV} \equiv F_{UV}/F_{X}$, which we fix at 10.  
 
As in Paper~1, our initial conditions are the solutions to the linearized versions of 
equations (1)-(3) (neglecting $\mathbf{f}_{rad}$)
that describe the evolution of a thermally unstable isobaric condensation mode 
(eqs. 11-14 in Field 1965).  
The amplitude of the density perturbation was increased 
from $\delta \rho /\rho_{eq} = 5\times10^{-5}$ to $\delta \rho /\rho_{eq} = 0.1$ 
in order to shorten the duration of the initial phase of cloud evolution; 
this had no noticeable effect on subsequent evolution. 
We use periodic boundary conditions in order to track the acceleration of the 
cloud over many domain lengths (a domain length is $l_x = 3.11\times10^{10}~\rm{cm}$).   
Our simulations were performed on a uniform grid 
with resolution $N_x = 1024$ in 1D and $[N_x,N_y] = [2048,1024]$ in 2D.

The effects of a time-varying flux would be minimal if clouds evaporated back into the 
confining medium at a much higher rate than new clouds are created.
In an upcoming paper, we will present the results of a study dedicated to exploring a 
wide variety of initial conditions and cloud morphologies, all of which are suggestive 
of a scenario in which continuous cloud production 
can sustain a significant cloud mass fraction despite losses from evaporation.  
For the present study,  
this finding implies that a turbulent flow regime with qualitatively similar properties is 
reached nearly independent of the initial conditions.  
Hence, it suffices to simply consider the simulations from Paper~1 and run them 
for a longer time.   

\begin{figure*}
\includegraphics[width=0.98\textwidth]{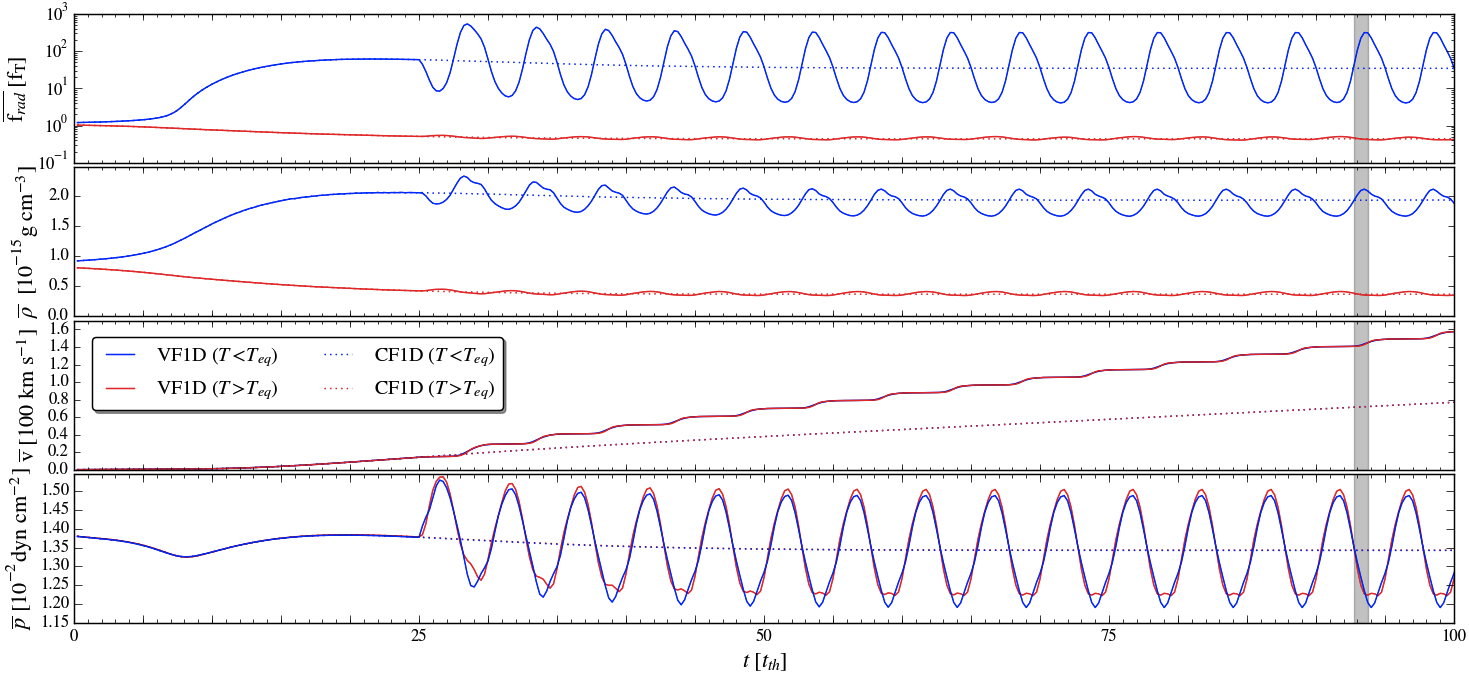}
\caption{Temporal properties of 1D simulations with (run VF1D; solid lines) 
and without (run CF1D; dotted lines) a time-varying flux.  
Red and blue colors denote averages over gas that is above and below, 
respectively, the equilibrium temperature $T_{eq} = 1.93\times10^5~\rm{K}$.  
In the top panel, $\mathrm{f}_{rad}$ is plotted in units of the force from 
Thomson scattering for gas with $T = T_{eq}$, 
namely $\mathrm{f_T} = \rho_{eq} \sigma_e (F_{UV} + F_{X})/c$ $[\rm{g~cm^{-2}s^{-2}}]$.
The grey region highlights the quarter cycle corresponding to the solutions plotted in Figure 2.  
The velocity panel shows that a 20\% variation in flux increases the net flow acceleration 
by about 240\%.}
\end{figure*}

\begin{figure}
\centering
\includegraphics[width=0.36\textwidth]{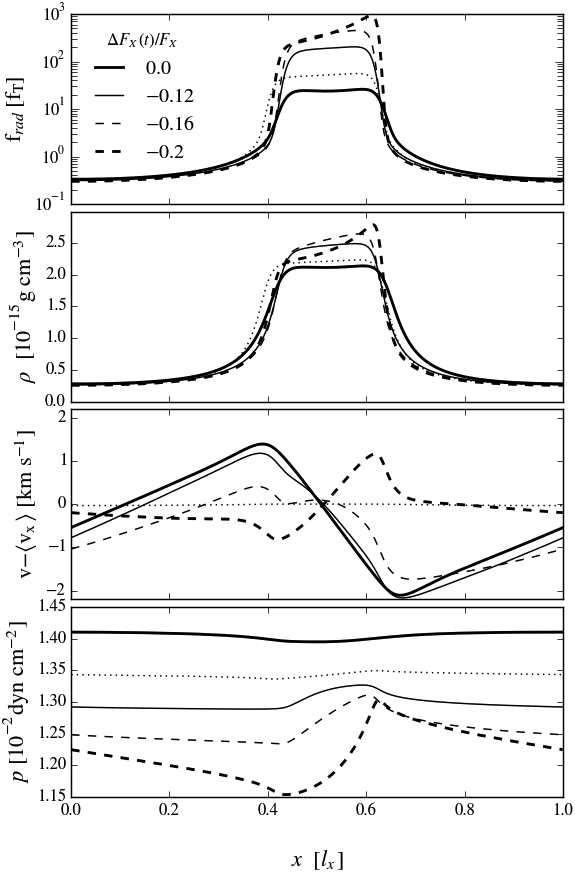} 
\caption{Spatial profiles in the comoving frame of the cloud for the quarter cycle 
(highlighted in grey in Figure 1) when the variable flux transitions from its equilibrium 
($\Delta F_X = 0$; thick solid line) to its minimum 
($\Delta F_X = -0.2\, F_{X}$; thick dashed line) value.  
The thin solid and dashed lines are intermediate profiles with 
values of $\Delta F_X/F_{X}$ shown in the legend.  
The dotted profiles are the solutions for (constant flux) run CF1D.
See the caption of Figure 1 for the definition of $\mathrm{f_T}$.}
\end{figure}

\begin{figure*}
\includegraphics[width=0.98\textwidth]{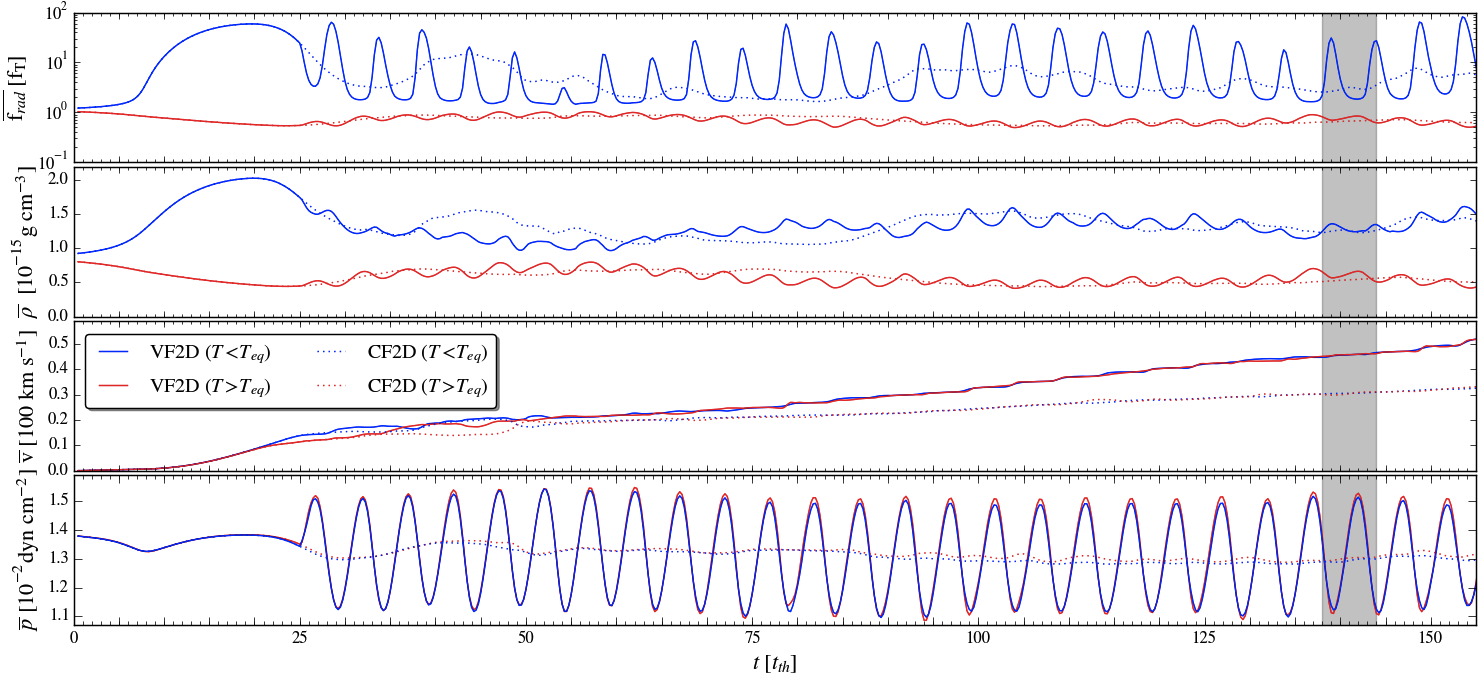}
\caption{Same as Figure 1, but for 2D simulations run for 155 thermal times.  
The grey region now highlights a duration spanning 1.25 cycles; 
snapshots of each quarter cycle are plotted in Figure 4.  
Once again, the velocity panel shows that a 20\% variation in flux 
increases the net flow acceleration by about 240\%.}  
\end{figure*}

\section{Results}
\label{sec:results}
We ran over 20 simulations to explore the parameter space 
($A_X$ and $\omega_X$) introduced by the time-varying flux.
Typical rms variability amplitudes observed in reverberation mapping 
campaigns are $\lesssim 20\%$ (e.g., De Rosa et al. 2015).  
For amplitudes this small, 
we expect there to be a limited range of periods 
$t_X = 1/\omega_X$ that can significantly
affect cloud acceleration.  If the flux varies rapidly, such that $t_X \ll t_{\rm{th}}$, 
the gas will not have time to respond.  The opposite regime with 
$t_X \gg t_{\rm{th}}$ would likely be very inefficient, as
our results show that each low flux state provides a gentle `kick' to the cloud.  

Here we present results for two simulations carried out in both 1D and 2D.  
Runs CF1D and CF2D (CF for `constant flux') are identical to the fiducial 
runs from Paper~1;   
VF1D and VF2D (VF for `variable flux') are new simulations 
and differ only by the introduction of $\Delta F_X(t)$.  
We adopted $t_X = 5\, t_{\rm{th}} \approx 0.35$ days.  
We do not introduce this time-varying flux until the cloud has fully formed.  
In practice, we set 
$A_X = 0$ if $t < 25\,t_{\rm{th}}$ and $A_X = 0.2$ for $t \ge 25\,t_{\rm{th}}$.

\subsection{1D Simulations}
In Paper~1 we showed that equations (1)-(3) reach a simple steady state solution 
in 1D when the flux is constant.  
Here the 1D solutions are much more complex and naturally time-dependent, 
but they are perfectly cyclic in that profiles of the solution at late times satisfy 
$q(x,t) = q(x,t + t_X) + q_0$, where $q$ is any variable and $q_0$ is a constant.  
This property is most easily shown by plotting spatially averaged quantities over time, 
as shown in Figure 1.  Here the instantaneous values of the radiation force 
and primitive variables are averaged over the hot gas ($T>T_{eq}$; red curves) 
and cold gas ($T<T_{eq}$; blue curves).  
Since the density and pressure averages are bounded by the same values for 
$t \gtrsim 40\,t_{th}$, we have $q_0 = 0$ for $\rho$ and $p$, 
indicating that a stable configuration permitting cyclic episodes of enhanced 
cloud acceleration has been reached.  
For comparison, the dotted lines in Figure 1 are these same quantities for run CF1D.  

Comparing red and blue curves in Figure 1,
 we see that our basic expectations are confirmed: 
low flux states lead to decreases in temperature, 
which for the cold gas results in accompanying increases in density 
(due to the tendency to maintain pressure equilibrium) 
as well as corresponding increases in the radiation force.   
At $t = 25\,t_{\rm{th}}$, the flux oscillations commence, 
beginning with a high state from $t = 25.0 - 27.5 \,t_{\rm{th}}$.  
The resulting increase in temperature is reflected by the initial rise in 
pressure in the bottom panel, and the cold and hot gas pressures are closely in sync, 
explaining the initial drop in the density of cold gas.  
(The much smaller rise in the density of the hot gas is slightly delayed, 
indicating that this is a hydrodynamic response to the cold gas.)  
The `kick' imparted by the radiation force happens in the next quarter cycle from 
$t = 27.5 - 28.75 \,t_{\rm{th}}$.  
The grey shaded region highlights this interval at a later time, 
showing that the slope of the velocity is steepest during this quarter cycle.   
 
In Figure 2 we investigate the detailed dynamics of the cloud during this 
acceleration phase of the cycle.  The two thick lines are profiles of the solution 
corresponding to times at the boundaries of the shaded region in Figure 1.   
The top panel shows that a 20\% variation in flux leads to more than an order of 
magnitude increase in the radiation force for run VF1D compared to run CF1D 
during this phase of the cycle. 
The cloud responds to this force by dramatically altering its configuration.  
For the cloud to remain in near pressure equilibrium with the hot medium 
and yet cool (reflected by the overall decrease in pressure in the bottom panel), 
its density must increase.
This in turn can only occur through a bulk transfer of mass from the hot medium, 
hence the prominent positive and negative peaks on the velocity profile, 
which are maintained for about a third of the quarter cycle.  
This advective mass transport though both interfaces of the cloud suddenly 
becomes much weaker through the left interface (see the thin dashed line), 
leading to a marked density increase of the leading edge of the cloud.  
Meanwhile, the pressure gradient in the core has progressively steepened, 
indicative of the increasing drag as the radiation force increases.  

The final velocity profile (thick dashed line) indicates that the bulk mass transfer 
opposes continued growth and will instead lead to a net expansion of the cloud.  
In the next quarter cycle (not shown, but see Figure 1), the cloud density drops 
back to its starting level in Figure 2, and then continues to further decrease as 
the flux increases, causing the temperature to rise and the cloud to expand. 
The term representing $p\,dV$ work, $p \nabla \cdot \mathbf{v}$, 
is critical for mediating the transition between high and low flux states.   
However, in the steady state $p\,dV$ work does not play a role since 
$\nabla \cdot \mathbf{v} = -\rho^{-1}D\rho/Dt$, and $D\rho/Dt$ reaches 0 for run CF1D.  
When time-varying radiation forces are involved, this term is important.    

\begin{figure}
\centering
\includegraphics[width=0.4\textwidth]{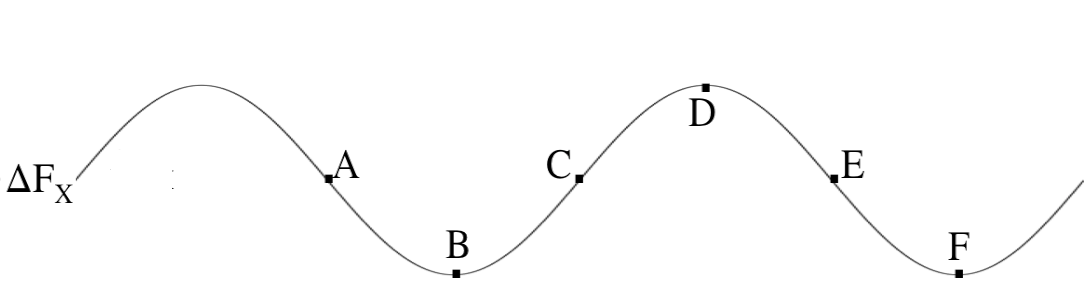}
\includegraphics[width=0.45\textwidth]{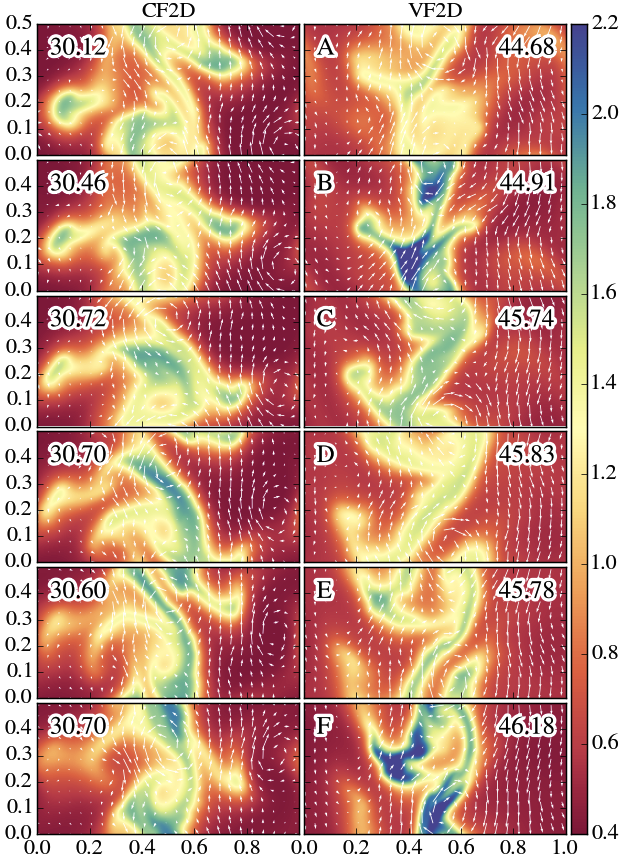}
\caption{Density maps of our 2D runs 
(colorbar values are in units of $10^{-15}\rm{g~cm^{-3}}$).  
Panels A-F on the right show snapshots of run VF2D every quarter cycle 
of the flux oscillation, as sketched above.  
The left maps are for run CF2D at the same times.  
Velocity vectors are overlaid after subtracting $\langle \mathrm{v}_x \rangle$, 
the mass-weighted mean of $\mathrm{v}_x$.  
This value is displayed on every panel in $\rm{km~s^{-1}}$ 
to show the enhanced acceleration during low flux states.
(The mass-weighting allows $\langle \mathrm{v}_x \rangle$ to decrease.)
}
\end{figure}

\subsection{2D Simulations}
The setup for our 2D simulations is the same as in Paper~1: 
we arrange for a plane parallel cloud (a slab) to be formed by making the magnitude
of the density perturbation in the $y$-direction 2 orders of magnitude smaller than that
of the $x$-direction: $(\delta \rho)_y /\rho_{eq} = 10^{-3}$.  This perturbation triggers
the Rayleigh-Taylor instability, as the core of the slab is slightly `heavier' than its 
surroundings under the effective gravity of the radiation force, and it `falls' into the 
hot medium.  At $t = 25\,t_{\rm{th}}$ when we apply the variable flux, the Rayleigh-Taylor
plume is fully developed 
(and similar to the configuration shown in the top right panel of Figure 4 in Paper~1).

Figure 3 is the 2D counterpart to Figure 1.
A two-phase medium clearly exists for the duration of the run, 
i.e. evaporation does not dominate new cloud production.  
While the generation of turbulence prevents an orderly cyclic solution, 
the behavior is qualitatively the same as in 1D.  
There are substantial quantitative differences: 
the average density of the cold gas never exceeds
its maximum at $t \approx 20\,t_{\rm{th}}$ 
and the radiation force is correspondingly weaker; 
hence, the velocities are greatly reduced.  
Importantly, however, the ratio of the net accelerations for runs VF2D and CF2D is 
approximately the same as that for runs VF1D and CF1D --- a factor of 2.4.  
This implies that the effect of a variable flux is quite robust. 

The right panels in Figure 4 show density maps of run VF2D for five consecutive, 
quarter-cycle flux states (denoted A-F in the sine-wave sketch), 
while the left panels are the corresponding density maps for run CF2D.   
During both of the minimum flux states (panels B and F), 
the density is noticeably increased, 
as hydrodynamic effects akin to those depicted in Figure 2 are taking place.  
Velocity arrows are overlaid in the comoving frame in order to
portray the local velocity field; 
they reveal the pronounced \emph{vortical} motions of the clouds, 
indicative of the large amount of vorticity in the flow.  
The mass-weighted mean velocity, $\langle \mathrm{v}_x \rangle$ (in $\rm{km~s^{-1}}$), 
is displayed on the corner of each panel in order to judge the net acceleration.  
The main noticeable effect is that during the low flux states, 
there are large increases in velocity compared to run CF2D.  

\section{Discussion and Conclusions}
As part of our ongoing effort to understand the dynamics of BLR clouds from
first principles, we have investigated the dynamical response 
of a two-phase medium to a time-varying flux.
Our main result is that small flux oscillations ($\Delta F_X = 20\%$) 
can lead to large changes in the net acceleration (240\%), 
even in 2D where the flow becomes highly turbulent.
The physics of this process is cleanly revealed in 1D, 
where a cyclic solution was found.
During every low flux state, the gas cools, 
allowing additional lines to appear that more strongly couples 
the gas to the radiation field and further accelerates the cloud via line driving.

Crucially, gas pressure effects are very important in mediating the
transition between flux states, 
thereby permitting the density to respond to the changes in temperature.    
Simply lowering the flux to its minimum value and then holding it fixed 
does not lead to an increase in acceleration; 
an explicit calculation revealed that in both 1D and 2D the final velocities 
were actually 4\% smaller than those for runs CF1D and CF2D.  
In other words, a time-varying flux leads to a gas pressure dominated, 
time-dependent solution that is qualitatively different from a constant flux solution.  

This finding may have interesting observational consequences and important 
implications for photoionization modeling efforts.  
For example, the responsivity of the BLR gas (Krolik et al. 1991; Peterson 1993) 
is a central quantity in reverberation mapping (RM) and transfer functions 
have been shown to be sensitive to how this quantity scales with radius (Goad et al. 1993).  
A negative responsivity, 
i.e. a decrease in line emission in response to an increase in the continuum flux, 
is readily interpreted as emission from optically thin clouds (Sparke 1993) and 
has been seen in X-ray RM observations (McHardy et al. 2007; Fabian et al. 2009).   
Photoionization models have difficulty accounting for this effect and simultaneously 
reproducing the observed line strengths 
(Shields et al. 1995; see also Snedden \& Gaskell 2007).  
The hydrodynamic effects associated with our present solutions 
--- significant decreases in density accompanying high flux states --- 
would naturally be expected to give rise to a negative responsivity, 
although they would complicate the analysis of time-delays in RM 
as they are a nonlinear response (e.g., Skielboe et al. 2015).  
To resolve the issue with line strengths, 
photoionization models may need to incorporate results from time-dependent hydrodynamics.

Our 2D simulations suggest that the long-term evolution of a two-phase medium in the BLR
is a highly turbulent flow that is conducive to continuous cloud production.  
The chaotic state is a consequence of vorticity generation, 
as neither of the two requirements to conserve vorticity 
(namely, that the radiation force be derivable from a potential and that the flow be barotropic) 
are close to being met.  
Once disrupted cloud fragments become small enough that their dimensions approach 
the length scale for conduction, i.e. the Field length $\lambda_F$ (Begelman \& McKee 1990), 
they will be subject to evaporation on a thermal timescale (Cowie \& McKee 1977).
This can be seen taking place in Figure 4, 
yet we find that cloud production can be maintained 
because the turbulence supplies perturbations that continually trigger the thermal instability.  
However, can this turbulent flow regime permit clouds to be accelerated to the 
velocities inferred from broad emission lines?  
 
 We can address this question if our simulations are representative of the
 local dynamics in a global simulation, 
 which should be the case for length scales over which the flux
 does not falloff substantially, say by more than 10\%.  
 This acceleration zone is $\Delta r \approx 0.05\, r_0$, 
 where $r_0$ is the distance where the cloud is formed.  
 For a $10^8~M_{\odot}$ black hole and 
 luminosity $L_X = 0.1\,L_{\rm{Edd}}$, we have $r_0 = \sqrt{L_X/4\pi F_{X}} = 44~\rm{ld}$.  
Assuming the cloud accelerates from rest, 
the velocity obtained is 
$\mathrm{v}_f = \sqrt{2\langle a\rangle \Delta r} \approx 2800~\rm{km~s^{-1}}$ for $\Delta r= 2.2~\rm{ld}$ 
and the average acceleration of run VF2D, $\langle a\rangle = 6.5~\rm{cm~s^{-2}}$.  
(We neglected gravity, as its acceleration is only $-1.0~\rm{cm~s^{-2}}$ at $r_0$.)  
This highly supersonic speed is reached in about 500 days, 
which may further complicate RM predictions (e.g., Waters et al. 2016),  
since it implies dynamical changes on timescales comparable to the duration of observational campaigns.
Nevertheless, it appears we are close to having a comprehensive theory of cloud formation and acceleration 
that can account for the high velocity BLR gas.   

\section*{Acknowledgments}
This work was supported by NASA under ATP grant NNX14AK44G.  
Our simulations were performed on the Eureka cluster at UNLV's 
National Supercomputing Center for Energy and the Environment.

\end{document}